\documentclass[preprint,aps, preprintnumbers,
nofootinbib,superscriptaddress]{revtex4}
\usepackage[dvips]{graphics}

\newcommand{\gsim}[2]{
\setlength{\unitlength}{12pt}
\begin{picture}(1.4,1.)
\put(.7,-0.3){\makebox(0.0,1.)[t]{$>$}}
\put(.7,-0.3){\makebox(0.0,1.)[b]{$\sim$}}
\end{picture}#2}

\begin{document}

\title{Constraining interacting dark energy models with flux
destabilization}

\author{Raul Horvat\footnote{Electronic mail address: horvat@lei3.irb.hr}}
\affiliation{Rudjer Bo\v{s}kovi\'{c} Institute,
         P.O.B. 180, 10002 Zagreb, Croatia}

\author{Diego Pav\'{o}n\footnote{Electronic mail address: diego.pavon@uab.es}}
\affiliation{Departamento de F\'{i}sica, Universidad Aut\'{o}noma
de Barcelona, 08193 Bellaterra (Barcelona), Spain}

\begin{abstract}
A destabilization in the transfer energy flux from the vacuum to
radiation, for two vacuum decay laws relevant to the dark energy
problem, is analyzed using the Landau-Lifshitz fluctuation
hydrodynamic theory. Assuming thermal (or near thermal)
equilibrium between the vacuum and radiation, at the earliest
epoch of the Universe expansion, we show that the law due to
renormalization-group running of the cosmological constant term,
with parameters chosen not to spoil the primordial nucleosynthesis
scenario, does soon drive the flux to fluctuate beyond its
statistical average value thereby distorting the cosmic background
radiation spectrum beyond observational limits. While the law
coming from the saturated holographic dark energy does not lead
the flux to wildly fluctuate, a more realistic non--saturated form
shows again such anomalous behavior.
\end{abstract}


\maketitle

Oftentimes, the present state of accelerated expansion of the
Universe is related to some mysterious dark energy sector. This is
linked to the longstanding cosmological-constant (CC) problem
\cite{nobb}, by adding to it two distinct (but connected)
difficulties: (i) the ``new'' CC problem, a puzzle of why the CC
is small but non-zero, and (ii) the ``cosmic coincidence"/``why
now?" problem \cite{paul}, a puzzle concerning the current near
coincidence of the CC energy density, $\rho_{\Lambda }$, with that
of matter despite they scale at different rates with expansion. It
seems today that a new aspect in dealing with the CC problems lies
in the landscape of string theory \cite{bousso1}, though making
predictions in such a theory constitutes an enormous challenge
\cite{bousso2}.

Long before the  ``environmental" variable CC approach of the
string theory landscape it was noticed that, up to some extent, a
traditional running of the CC can ameliorate the fine-tuning
problems inherent to the CC by providing a viable mechanism to
efficiently relax it from a very high value at the early Universe
to its current tiny value. It was subsequently noted that some of
the running CC models could successfully mimic the popular
quintessence models as well as shed some light on the coincidence
problem, thus becoming viable cosmological models of dark energy
of the Universe. Arguably, the most appealing amongst them are
those whose laws for the CC running can be inferred from some
underlying physical theory. So, some of the most attractive dark
energy models involve the CC running laws derived from quantum
theory of particle fields on the classical gravitational
background \cite{sola1}, quantum gravity \cite{bonanno} and
gravitational holography \cite{cohen}. A comprehensive list of
phenomenological laws for the CC variation under consideration
well before the discovery of dark energy, can be found in
\cite{overduin}.

Barring a time-dependent gravitational coupling \cite{stef, gub1,
bonanno} or going over some scalar-tensor theory
\cite{scalar-tensor}, the simplest way to achieve the infrared
(IR) screening of the CC is through its decay into matter and/or
radiation. The interesting possibility in this context, put
forward long ago, was to consider the cosmological vacuum decay
into radiation as a measure of the temperature of the vacuum
\cite{mauricio}. Thus, if the interactions between the two
components at the earliest moments of the Universe expansion were
fast enough to bring them in thermal or near thermal equilibrium,
then both -the vacuum and radiation- would share a common
temperature for, at least some time, during the expansion.
Interestingly,  such a scenario would preclude a thermal
equilibrium between the vacuum and the event horizon (which
naturally occurs in these scenarios for most of the history of the
Universe), for much of the time except near the Planck time.
Indeed, since the temperature of the horizon as given by the
Hubble parameter $H$, and the temperature of the vacuum/radiation
$T$, scale differently with the expansion, they may therefore
coincide (during a radiation dominated epoch), $H \simeq T$, only
when $T \simeq M_{Pl}$.

At a macroscopic level, the transfer of energy from vacuum to
radiation (and vice versa) is governed  by the continuity equation
\\
\begin{equation}
\dot{\rho}_{\Lambda } + \dot{\rho}_R  + 4H\rho_R = 0 \;,
\label{continuity}
\end{equation}
\\
with $\rho_R $ being the radiation energy. Once the equilibrium
(or near equilibrium) between the sub-systems vacuum/radiation
gets established it will remain so provided the heat capacity of
the whole system is positive-definite. Since the radiation heat
capacity is necessarily positive, this amounts to having the heat
capacity of vacuum
\\
\begin{equation}
C_{\Lambda } = T \left( \frac{\partial S_{\Lambda }}{\partial T}
\right )_{V} \;
\label{heatc}
\end{equation}
\\
positive, where $S_{\Lambda }$ represents the entropy of a
variable CC. (Admittedly, there is some ambiguity when taking the
partial derivative in Eq. (\ref{heatc}) as the volume should be
kept constant but the latter depends on T in an expanding
Universe).

Our target in the present paper is to study the fluctuations of
the flux $\dot{\rho}_{\Lambda }$ entering Eq. (\ref{continuity})
around its statistical average value, from the laws emerging from
the RG-running and gravitational holography, taking the
macroscopic criterion that $C_{\Lambda} > 0$ as a consistency
condition. To fulfill this aim we shall employ the well known
Landau-Lifshitz (LL) fluctuation hydrodynamic theory \cite{lev},
which applies to equilibrium and nonequilibrium classical
statistical theory \cite{miguel}. Our particular emphasis will be
on finding a scale dependence of these fluctuations, with a
stabilization criterion that the root mean square of the
fluctuations must never exceed the average value of the
corresponding flux.

According to Landau and Lifshitz,  if the flux $\dot{y}_{i}$ of a
given thermodynamic quantity, which evolves in a generic
dissipative process, is governed by $\dot{y}_{i} = \Sigma_{j} \,
\Gamma_{ij}\, Y_{j} + \, \delta\dot{y}_{i}$ and the entropy rate
obeys $\dot{S} = \Sigma_{i}\, (\pm Y_{i}\, \dot{y}_{i})$, then the
second moments of the fluctuations of the fluxes are given by
$<\delta\dot{y}_{i}\, \delta\dot{y}_{j}> = (\Gamma_{ij} +
\Gamma_{ji})\, \delta_{ij} \, \delta(t_{i}- t_{j})$. The angular
brackets denote statistical average with respect to the reference
state (i.e., $ \Sigma_{j} \, \Gamma_{ij}\, Y_{j}$ which is
supposed to be steady or quasi-steady and constitutes the
systematic part of the flux), and the fluctuations
$\delta\dot{y}_{i}$ are considered spontaneous departures from
that  state, thus $<\delta\dot{y}_{i}>$ vanishes identically. The
quantities $\Gamma_{ij}$ and $Y_{i}$ stand for the
phenomenological transport coefficients and the thermodynamic
force conjugate to the flux $\dot{y}_{i}$, respectively. In the
expression for $\dot{S}$ the minus sign must be taken when the
product $Y_{i}\, \dot{y}_{i}$ is negative, otherwise the plus sign
should be considered. This theory has been successfully employed
to constrain models for the decay of the cosmological constant
into radiation and/or matter \cite{L-decay} as well as in the
analysis of second order nonequilibrium phase transitions in
isolated black holes \cite{bh}.

The LL theory when applied to the decay of a variable CC includes
just a single flux, $\dot{\rho}_{\Lambda }(t)$,  governed by
\footnote{The second law of thermodynamics implies the presence of
particle creation and therefore the only flux to be considered is
$\dot{\rho}_{\Lambda }$.}
\\
\begin{equation}
\dot{\rho}_{\Lambda } = \Gamma \; Y + \delta \dot{\rho}_{\Lambda }
\; , \label{dotrholambda}
\end{equation}
\\
where the   fluctuations, $\delta \dot{\rho}_{\Lambda }$, coming
from the decay of the vacuum are assumed to be {\it Gaussian},
random fluctuations, with uncorrelated Fourier modes due to
statistical homogeneity and isotropy. The thermodynamic force
conjugate to the flux $\dot{\rho}_{\Lambda }$ follows from
combining the entropy production rate
\\
\begin{equation}
\dot{S}_{\Lambda } = Y \dot{\rho}_{\Lambda } \;, \label{dotS}
\end{equation}
\\
with Eq. (\ref{dotrholambda}). Finally, the second moment (i.e.,
the root mean square) of the fluctuations of the flux is given by
\\
\begin{equation}
<\delta \dot{\rho}_{\Lambda }(t_i ) \, \delta \dot{\rho}_{\Lambda
}(t_j )> = 2\,  \Gamma \, \delta (t_i -t_j ) \;.
\label{rms}
\end{equation}

For the models explored below we have that
\begin{equation}
\dot{\rho}_{\Lambda } \sim \rho_{\Lambda }H \, .
\label{similar}
\end{equation}
Near $t = t_{Pl}$, we are in the realm of $a \sim t^{\alpha }$
cosmologies (see below) thereby both terms on the right hand side
of Eq.(\ref{similar}) vary less and less faster with time. In
addition, for the RG-running model, both $\rho_{\Lambda }$ and $H$
will approach a constant value at late times but well before the
present moment (see below). Thus, the approximate steady-state
regime can be maintained even for $t \rightarrow \infty $,
provided $C_{\Lambda}
> 0$.

Using the above sketched LL theory for the fluctuations of the
fluxes, we aim to study  their behavior for the approach to the CC
problem based on the RG \cite{sola1}. The RG is a conventional
theoretical tool for investigating quantum effects and the scale
dependence of a certain quantity. From the viewpoint of quantum
theory of matter fields in curved space \cite{birell, buchbinder},
the renormalizability of the theory forces the vacuum action to
contain the CC term as well as fourth derivative terms. Then, the
CC term is viewed as a parameter subject to RG running and
therefore it is expected to run with the RG scale, usually
identified with an expansion quantity evolving smoothly enough to
comply with the cosmological data. In such theories, therefore,
even a ``true" CC cannot be set to any definite fixed value
(including zero) owing to the RG running effects. It may be
surprising, however, that the time-dependence of the CC may be due
to quantum effects from the RG, considering the familiar quadratic
decoupling of heavy matter fields at low energy.\footnote{Strictly
speaking, the quadratic form of decoupling can be proved in a
rigorous way only for higher derivative terms of the vacuum
action, but not for the CC term itself \cite{gorbar}. Yet,
usually, the same is assumed for the CC term.} The reason for this
result \cite{gub2} lies in the high dimensionality $({\it mass}^4
)$ of the scaling quantity $\rho_{\Lambda }$, with the outcome
that the more massive a field is, the more dominant the role it
plays in the running -irrespective of the scale. Consequently, the
running becomes stronger than logarithmic, thus providing  a
dynamical, efficient relaxation mechanism for the CC to go down
the tiny value we observe today. Further, the above scenario for
the CC running, with the choice for the RG scale $\mu = H$, may
also furnish a viable cosmological model of dark energy
\cite{sola2, sola3}. The strongest phenomenological constraints on
the RG model (within a framework where the running of the CC goes
at the expense of the energy transfer between vacuum and dark
matter) have been obtained recently by analyzing density
perturbations for the running CC \cite{sola4} and considering the
validity of the generalized second law for the running CC scenario
\cite{horvat1}.

The solutions for the RG-running vacuum decay into radiation for
flat space read \cite{sola2, sola3},
\\
\begin{equation}
\rho_R = \rho_{R0} a^{-4(1-\nu )} \;,
\label{rhoR}
\end{equation}
\\
\begin{equation}
\rho_{\Lambda} = \left(\rho_{\Lambda 0} - \frac{\nu }{1 - \nu }\,
\rho_{R0}\right) + \rho_{R0} \, \frac{\nu }{1 - \nu } a^{-4(1-\nu
)} \; ,
\label{rholambda}
\end{equation}
\\
\begin{equation}
H^2 = \frac{8 \pi }{3} M_{Pl}^{-2} \left[\left(\rho_{\Lambda 0} -
\frac{\nu }{1 - \nu }\rho_{R0}\right) + \rho_{R0} \frac{1}{1 - \nu
} a^{-4(1-\nu )}\right] \;,
\label{Hsquare}
\end{equation}
\\
where $\nu = \frac{\sigma }{12 \pi } \frac{M^2 }{M_{Pl}^2 }$ is a
dimensionless mass parameter driving the RG running. Here $M$
represents an additive mass contribution of all virtual massive
particles, $\sigma = \pm 1$ depending on whether the highest-mass
particle is a boson or a fermion, and the zero subscript  denotes
present-day value. Accordingly, $|\nu | \sim 10^{-2}$ would signal
the existence of a particle  with Planck mass (or the existence of
somewhat less massive particles with large multiplicities); $|\nu
| \sim 1$ would indicate the existence of a particle with
trans-planckian mass; $|\nu |  \sim 10^{-6}$ would mean the
existence of a particle with mass at the GUT-scale, whereas much
smaller values of $|\nu |$ would imply an approximate cancellation
between bosonic and fermionic degrees of freedom.

For adiabatic vacuum decays one expects the Stefan law to be
preserved (in the following we shall omit numerical factors since
it is only the functional dependence that matters). Using Eq.
(\ref{rhoR}) and bearing in mind that $\rho_{R} \propto T^{4}$, we
get
\\
\begin{equation}
T \sim a^{-1 + \nu } \; ,
\label{temperature}
\end{equation}
\\
and since $T$ is not expected to increase with expansion we infer
that $0 <\nu < 1$.

The entropy production associated to the CC decay is given by
Gibb's equation with the chemical potential set to zero
\\
\begin{equation}
\dot{S}_{\Lambda} = \frac{1}{T} V \dot{\rho}_{\Lambda} \;,
\label{dotSlambda}
\end{equation}
\\
where $V \sim a^3 $. From Eqs.  (11) and (4) we obtain
\\
\begin{equation}
Y = \frac{V}{T} \sim a^{4 - \nu } \, .
\label{kapitalY}
\end{equation}
\\
A combination of $\dot{\rho}_{\Lambda} \sim a^{-4(1-\nu)} H$, from
Eq. (\ref{rholambda}), with Eq. (\ref{dotrholambda}) yields
\\
\begin{equation}
\Gamma \sim a^{5\nu -8} H \,  .
\label{Gamma}
\end{equation}
\\
We then determine the dimensionless ratio between the second
moment of the fluctuations and the (squared) flux
\\
\begin{equation}
\eta \equiv \frac{<\delta \dot{\rho}_{\Lambda} \delta
\dot{\rho}_{\Lambda }>}{\dot{\rho}_{\Lambda} \dot{\rho}_{\Lambda}}
\sim \frac{\Gamma }{(\dot{\rho}_{\Lambda})^2 } =  a^{-3\nu} H^{-1}
\;. \label{eta}
\end{equation}
\\

Notice that the scaling dependence of the ratio $\eta$ is crucial.
Should $\eta$ increase with expansion, sooner or later the
fluctuations of the flux would become larger than the statistical
average value of the flux, signaling destabilization -i.e., the
flux would exhibit an erratic, unphysical, behavior. This would
mean that there is not longer guarantee that the fluctuations of
the flux preserve the equilibrium relations of the adiabatic decay
of the vacuum, especially the Stefan law \cite{lima}. Further,
this would seriously upset the black--body spectrum of the cosmic
microwave background (CMB) beyond the limits allowed by
observation. The opposite instance ($\eta$ decreasing with
expansion), corresponds to the usual, stable condition.

Using $H \sim a^{-2(1 -\nu )}$ from Eq. (\ref{Hsquare}) for $a \ll
1$, we get $\eta \sim a^{-5\nu + 2}$. Then, by requiring that
$\eta$ does not increase with expansion, we obtain
\\
\begin{equation}
\nu \geq \frac{2}{5} \, .
\label{2/5}
\end{equation}

On the other hand, for $a \gg 1$ we have that $H \sim$ constant
whereby $\nu > 0$ follows. From Eqs. (\ref{heatc}) and
(\ref{dotSlambda}) we get
\\
\begin{equation}
C_{\Lambda} \sim \frac{4\nu}{1-\nu}\, a^{3\nu }\, ,
\label{Clambda}
\end{equation}
\\
whence for the allowed range of values for $\nu$ the heat capacity
of the vacuum results positive-definite and thermal equilibrium
between vacuum and radiation is to hold for ever.

Before proceeding, it is expedient to check whether the systematic
part of the flux  is quasi-steady. It will be whenever the time
scale for the vacuum to decay into radiation is larger than the
expansion time (i.e., $\rho_{\Lambda}/\mid\dot{\rho}_{\Lambda}\mid
\gsim \, H^{-1}$). Noting that, as follows from Eq.
(\ref{rholambda}), the systematic part is $\dot{\rho}_{\Lambda} =
- 4 \, \rho_{R0} \, \nu \, a^{-4(1-\nu)}\, H$, it is readily seen
this is guaranteed for $\nu \gsim \, 3/4$.

Next, we compare our LL bound, given by (\ref{2/5}), with the
existing bounds on the RG model at any epoch. As is
observationally known, at the primordial nucleosynthesis time the
ratio $\rho_{\Lambda}/\rho_{R}$ did not exceed $0.05$
\cite{rachelbean}. Since
\\
\begin{equation}
\frac{\rho_{\Lambda }}{\rho_R } \simeq \frac{\nu }{1- \nu } \; ,
\label{ratio-rho-rho}
\end{equation}
\\
as follows from Eqs. (7) and (8) for $a \ll 1$, this sharply
contrasts with the LL bound (15).

The above formulae also apply in the case of a dynamical CC
scenario generically dubbed ``holographic dark energy" (HDE).
Originally derived  for zero-point energies \cite{cohen} as a
bound on $\rho_{\Lambda }$, the saturated form of the HDE is
usually written as \cite{li}
\\
\begin{equation}
\rho_{\Lambda } = \frac{3\, M_{Pl}^2}{8 \pi }\,  c^2 L^{-2} \, ,
\label{saturatedHDE}
\end{equation}
\\
where $L$ denotes the size of the region (providing an IR cutoff)
and $c^2$ is a dimensionless constant. This is a very important
concept since for $c^2 $ values of the order of unity, the HDE
model also provides  a very elegant solution of the ``old" CC
problem. Thanks to the relationship between the ultraviolet (UV),
$\rho_{\Lambda } \sim {\Lambda }^4 $, and the IR cutoff, the
holographic information is consistently encoded in the
conventional quantum field theory. The choice  $L = H^{-1}$ is
clearly the most natural and simple possibility \cite{horvat2,
pavon1, pavon2}. Then, with the aid of Friedman's equation we can
write
\\
\begin{equation}
c^2 = \frac{1}{1 + r_0 } \, , \label{csquare}
\end{equation}
where $r_0 =\rho_{R0}/\rho_{\Lambda0}$.

In order to connect the above formulas for the RG case to Eq.
(\ref{saturatedHDE}), we shall write $\rho_{\Lambda }$ from Eq.
(\ref{rholambda}) in a different fashion, namely,
\\
\begin{equation}
\rho_{\Lambda } = C_0 + C_2 H^2
\label{C+C}
\end{equation}
\\
with
\\
\begin{equation}
C_0 =  \rho_{{\Lambda }_0 } -\frac{3 \nu }{8 \pi } M_{Pl}^2
H_{0}^2 \, , \qquad \qquad  C_2  = C_0 + \frac{3 \nu }{8 \pi }
M_{Pl}^2 \, .
\label{C+Cwith}
\end{equation}
\\
The system of equations (\ref{C+C})-(\ref{C+Cwith}) is thus
equivalent to the (\ref{rhoR})-(\ref{Hsquare}) set. The HDE law
(\ref{saturatedHDE}), follows from (\ref{C+C})-(\ref{C+Cwith}) by
setting $C_0 =0$. Then, Eqs. (\ref{saturatedHDE}) and
(\ref{csquare}) are readily recovered -modulo, the obvious
identification $\nu = c^2 $. Now, as seen from Eq.
(\ref{csquare}), the bound from the LL theory, $c^2 \geq
\frac{2}{5}$, is respected since the ratio $r_0 $ is tiny today,
$\sim 10^{-5}$. Note that although the HDE law, Eq.
(\ref{saturatedHDE}) with (\ref{csquare}), describing the vacuum
decay, does intrinsically satisfy the LL bound, it does disturb
the big bang nucleosynthesis scenario by a wide margin. The main
problem is that the ratio $\rho_{R}/\rho_{\Lambda }$ stays frozen
during the whole cosmic expansion so that, for parameters driving
the accelerated expansion of the Universe at late times, a
transition to a radiation-dominated Universe is not feasible.
Thus, the model cannot be considered realistic.

A more realistic class of models, which do allow transitions
between the cosmological eras, is provided by the non-saturated
HDE concept \cite{pavon1, pavon2, horvat3}. The parameterization
is again given by Eqs. (\ref{saturatedHDE}) and (\ref{csquare}),
but with $c^2 $ promoted to a function of cosmic time whence the
ratio between the energy densities becomes a function of time. A
criterion for a realistic non-saturated HDE model is to saturate
the holographic bound asymptotically, $c^2(t \rightarrow \infty )
= 1$, while having $c^2 < 1$ in the radiation-dominated era. This type of
parametrization 
of $\rho_{\Lambda }$ in a non-saturation regime is particularly appealing since 
it reduces directly to (18), where again only the genuine IR cutoff shows 
up. It is
easy to find a particular model that does not comply with the LL
bound and the big bang nucleosynthesis bound simultaneously. It is
possible to find such a function $c^2 $, which does satisfy the
above criterion for a realistic non-saturated HDE model, so that
the RG law (\ref{C+C})-(\ref{C+Cwith}) become equivalent to the
law obtained from the non-saturated HDE. Indeed, the choice
\\
\begin{equation}
r(t) =r_0 \frac{a^{-4 + \epsilon }}{1 -\alpha r_0 (1 - a^{-4 +
\epsilon })\, (1 -\alpha + r_0)^{-1}} \, \label{r(t)}
\end{equation}
\\
with $\epsilon = 4 \alpha /(1 + r_0)$ and $\alpha \equiv C_2
H_{0}^2/\rho_{{\Lambda }_0 }$ reproduces a non-saturated HDE model
\\
\begin{equation}
\rho_{\Lambda } = \frac{3\, M_{Pl}^2}{8 \pi }\,  c^2(t) \, L^{-2}
\; , \label{nonsaturatedHDE}
\end{equation}
\\
with $c^2 = \frac{1}{1 + r(t)}$, equivalent to the RG-running law
given by Eqs. (\ref{C+C})-(\ref{C+Cwith}). As shown above, this
model does not satisfy the LL bound without seriously affecting
the big bang nucleosynthesis scenario.

In summary, two of the most popular and viable dark-energy models,
based on vacuum decaying laws, appear largely compromised by flux
destabilization. Our analysis relies on the assumption that, at
some time close to the Planck era, the vacuum was in thermal (or
near thermal) equilibrium  with radiation. We have shown that the
said equilibrium would persist at the time of  big bang
nucleosynthesis, where the vacuum is restricted to a tiny fraction
of the total energy density \cite{rachelbean}. If, in the
meantime, the fluctuations of the energy flux  become erratic
(i.e., $\eta > 1)$, then the radiation component will no longer
present a black--body spectrum and the CMB will get seriously
distorted. For a running CC scenario it has been shown that it is
not possible to simultaneously hinder the growth of fluctuations
(relative to the average value) and reduce the vacuum contribution
at nucleosynthesis' time to an acceptable level. For a saturated
HDE model, the flux remain under control at all times. However, if
the nucleosynthesis bound is fulfilled, then an accelerated
expansion  at late times cannot be achieved. The non--saturated
HDE model shows identical anomaly as the RG model.

We may conclude by saying that either a thermal equilibrium
between vacuum and radiation did never occur or the dark energy
models here considered are in need of revision.

\acknowledgments{This work was partly supported by the Ministry of
Science, Education and Sport of the Republic of Croatia under
contract No. 098-0982887-2872, the Spanish Ministry of Education
and Science under Grant FIS 2006-12296-C02-01, and the
``Direcci\'{o} General de Recerca de Catalunya" under Grant 2005
SGR 00087.}

\end{document}